# Deflection of MEMS Based Sandwiched Cantilever Beam for Piezoelectric Actuation: Analysis of Lead Zirconate Titanate Piezoceramic Material

Iftekar Mahmud[1]

[1](Department of Mechanical and Aerospace Engineering, Missouri University of Science & Technology, USA)

*Abstract:*
*Background*: This paper investigates the deflection of micro- electro-mechanical system (MEMS) based sandwiched cantilever beam for piezoelectric actuation using eight of the popular lead zirconatetitanate (PZT)piezoceramic material. The simulation for the investigation was carried away in COMSOL multiphysicssoftware environment. For the investigation of a certain model, the lowest tip deflection is found to be 38.074 nm for PZT-8 and the highest tip deflection is found to be 82.965 nm for PZT-5H.
*Materials and Methods*: A 2 mm thick flexible foam core is sandwiched between two 8 mm thick aluminum layers in this model of sandwiched cantilever beam, which is 100 mm long. The gadget is replaced with a foam core with a 10 mm long piezoelectric actuator that is located between x=55 mm and x=65 mm. Along the global x-axis, the beam is aligned. Figure 1 depicts a visual representation of the geometry. Several assumptions are regarded the basic assumptions for the simulation to ease the formulation approach. And the boundary conditions are: 1) Structural mechanics: The beam is set at its surface x=0 and the other surfaces are free. 2) Electrostatics: A 20 V potential difference is applied between the piezoelectric domain's top and bottom surfaces. This creates a transverse shear strain by creating an electric field perpendicular to the poling direction (x-direction).
*Results*: The beam changes substantially when the piezoceramic material is changed. The tip deflection of the cantilever beam varies between 37 and 83 nanometers for different lead zirconate titanate piezoceramic materials. PZT-8 has the lowest deflection of 37.074 nm, whereas PZT-7A has the second lowest deflection of 40.423 nm.
*Conclusion:* The maximum tip deflection discovered is more than twice as high as the lowest tip deflection discovered. The deflection of a sandwich cantilever beam for piezoelectric actuation is clearly influenced by piezoceramic materials which can be used in increasing the efficiency of any mechanical device.

---

---

## I. Introduction

Piezoelectric actuation based on microelectromechanical systems (MEMS) is gaining appeal in a wide range of applications, from industrial machinery to small electronic devices. The main characteristic of piezoelectric actuation is that the piezoelectric ceramic material used in the piezoelectric actuator generates electrical energy when subjected to mechanical energy (piezoelectric effect) and mechanical energy when subjected to electrical energy (inverse piezoelectric effect). This sandwiched cantilever beam is required for the construction of big industrial and commercial buildings with sandwich paneled structures.

Sun and Zhang proposed the sandwich cantilever beam with shear induced bending mechanism in the field of piezoelectric actuation in 1995 [1]- [2]. Benjeddou et al. modeled a sandwich beam in 1997 [3]. This subject has been the subject of numerous studies. A sandwich plate with a piezoelectric core was investigated [4]. A numerical model is used to analyze piezoelectric smart composites [5]. For Piezoelectric Sandwich Plates, Mindlin-Type Finite Elements have been presented [6]. For Rectangular Sandwich Plates with Embedded Piezoelectric Shear Actuators, an Exact Solution to be used in the parabolic solar heater has been investigated [7]. Exact deflection solutions were investigated of beams with shear piezoelectric actuators used in natural fiber composite material [8]. The MEMS Based Piezoelectric Shear Actuated Beam was studied in terms of design and simulation for the use of condensed water recycling in any air conditioning unit. [9]. These above-mentioned applications [5-9] of the piezoelectric materials help to increase the efficiency of those devices. Modeled Micromechanical System Based Finite Element Piezoelectric Shear Actuated Beam was used to evaluate performance [10] [11].

The effect of eight popular Lead ZirconateTitanate (PZT) piezoceramic materials on the deflection of a MEMS-based sandwiched cantilever beam for piezoelectric actuation is demonstrated in this paper.

## II. Methods And Modeling

The modeling was done in the COMSOL software version 4.3 environment.

### A. Geometry

A 2 mm thick flexible foam core is sandwiched between two 8 mm thick aluminum layers in this model of sandwiched cantilever beam, which is 100 mm long. The gadget is replaced with a foam core with a 10 mm long piezoelectric actuator that is located between x=55 mm and x=65 mm. Along the global x-axis, the beam is aligned. Figure 1 depicts a visual representation of the geometry.

### B. Assumptions

Following assumptions are regarded the basic assumptions for the simulation to ease the formulation approach. Zhang and Sun were the leading proponents of these beliefs. [4].

1) In each layer, normal stress is believed to have dissipated.
2) The interfaces between neighboring layers are flawlessly adhered to one another.
3) Transverse displacements are equal in all layers.
4) The top and bottom layers are made out of traditional plates.
5) Inertia terms in rotary motion are believed to be minimal.
6) Transverse isotropy is postulated for piezoelectric materials.
7) The electrodes are entirely coated on two surfaces of the piezoelectric layers.

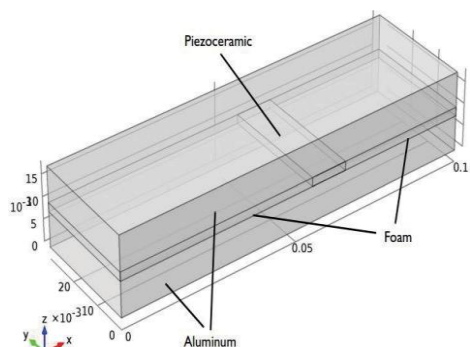

*Figure 1: 3D view of the sandwiched cantilever beam model*

### C. Boundary Conditions

1) Structural mechanics: The beam is set at its surface x=0 and the other surfaces are free.
2) Electrostatics: A 20 V potential difference is applied between the piezoelectric domain's top and bottom surfaces. This creates a transverse shear strain by creating an electric field perpendicular to the poling direction (x-direction).

### D. Material Properties

Properties of the used materials are illustrated in Table I.

*Table 1: MATERIAL PROPERTIES*

| Property | Aluminium | PZT-2 | PZT-4 | PZT-4D | PZT-5A | PZT-5H | PZT-5J | PZT-7A | PZT-8 | Foam |
|---|---|---|---|---|---|---|---|---|---|---|
| Density [q] (kg/☐³) | 2700 | 7600 | 7500 | 7600 | 7750 | 7500 | 7400 | 7700 | 7600 | 32 |
| Young's Modulus [E] (GPa) | 70 | — | — | — | — | — | — | — | — | 35.3 |
| Poisson's Ratio [v] | 0.35 | — | — | — | — | — | — | — | — | 0.383 |
| Relative Permittivity [$s_r$] | — | {990, 990, 450} | {1475, 1475, 1300} | {1610, 1610, 1450} | {1730, 1730, 1700} | {3130, 3130, 3400} | {2720, 2720, 2600} | {930, 930, 425} | {1290, 1290, 1000} | — |

*Deflection of MEMS Based Sandwiched Cantilever Beam for Piezoelectric Actuation: ..*

## III. Simulation

The heat study was conducted out using the finite element method in the COMSOL Multiphysics software environment. To finish the simulation, three phases were required: pre-processing, solver execution, and post-processing.

The first step was to decide which physics to use. Then it was time to start specifying geometry, materials, physics, meshing, and simulation computation one by one.

## IV. Results & Discussion

From Figure 2 to Figure 9, the simulation's outcomes are visibly depicted. The deflection is shown by the hue of the solid body. The more deflection the body represents, the redder it is, and the more violet it is, the less deflection it indicates. The angular deflection is measured in nanometers. It can be seen that the deflection of the beam changes substantially when the piezoceramic material is changed. The tip deflection of the cantilever beam varies between 37 and 83 nanometers for different lead zirconate titanate piezoceramic materials. PZT-8 has the lowest deflection of 37.074 nm, whereas PZT-7A has the second lowest deflection of 40.423 nm. PZT-2 has the third lowest deflection at 49.265 nm. PZT-5H has the largest deflection of 82.965 nm, whereas PZT-5J has the second highest deflection of 74.906 nm. PZT-5A has the third highest deflection at 65.322. As a result, different lead zirconatetitanatepiezoceramic materials have varying tip deflections.

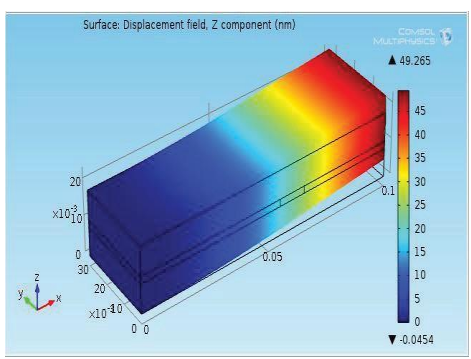

*Figure 2: Simulation result for PZT-2*

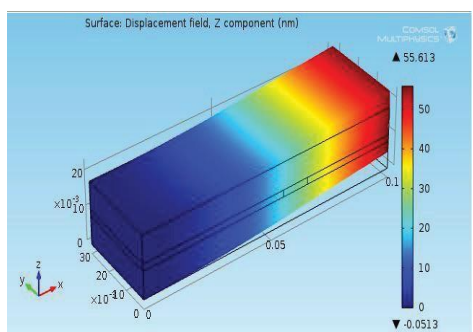

*Figure 3: Simulation result for PZT-4*

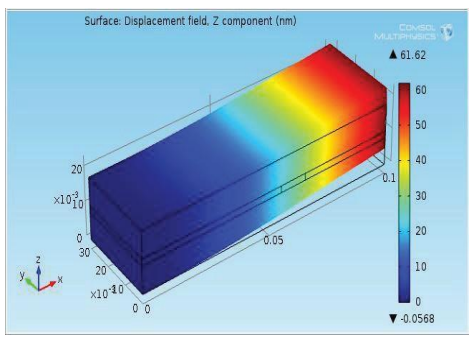

*Figure 4: Simulation result for PZT-4D*

DOI: 10.9790/1684-1805032226	www.iosrjournals.org	24 | Page

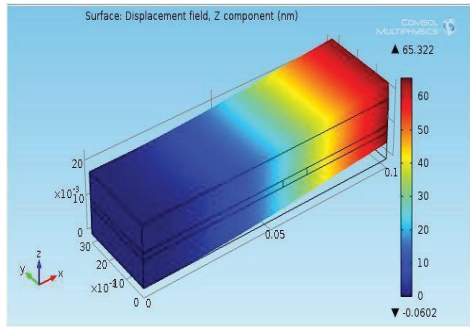

*Figure 5: Simulation result for PZT-5A*

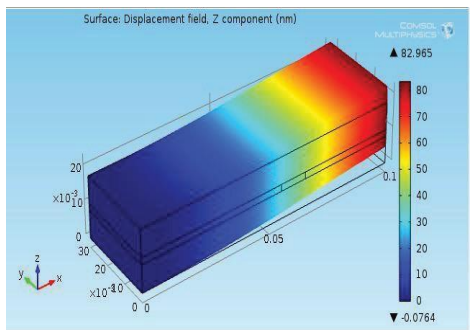

*Figure 6: Simulation result for PZT-5H*

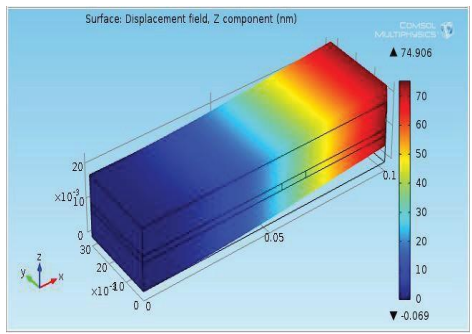

*Figure 7: Simulation result for PZT-5J*

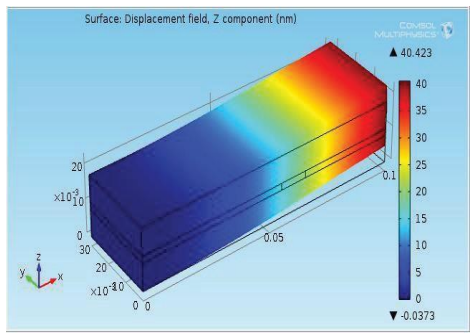

*Figure 8: Simulation result for PZT-7A*

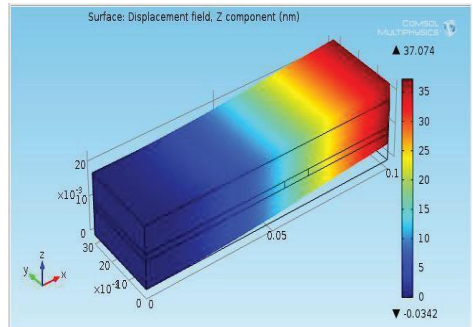

*Figure 9: Simulation result for PZT-8*

Table II summarizes the tip deflections determined from simulations for various lead zirconatetitanatepiezoceramic materials.

**Table 2:** TIP DEFLECTION FOR DIFFERENT LEAD ZIRCONATE TITANATE PIEZOCERAMIC MATERIAL

| Simulated Result | Piezoceramic Material | | | | | | | |
|---|---|---|---|---|---|---|---|---|
| | PZT-2 | PZT-4 | PZT-4D | PZT-5A | PZT-5H | PZT-5J | PZT-7A | PZT-8 |
| Tip Deflection | 49.265 | 55.613 | 61.62 | 65.322 | 82.965 | 74.906 | 40.423 | 37.074 |

## V. Conclusion

In this study, a MEMS-based sandwiched cantilever beam model for piezoelectric actuation was built in the COMSOL multiphysics virtual environment. Eight typical lead zirconatetitanatepiezoceramic materials were used to investigate beam deflection. The maximum tip deflection discovered is more than twice as high as the lowest tip deflection discovered. The deflection of a sandwich cantilever beam for piezoelectric actuation is clearly influenced by piezoceramic materials. Researchers can utilize this work to investigate the influence of a MEMS-based Sandwiched Cantilever Beam for Piezoelectric Actuation.